\documentclass[aps,pra,twocolumn,superscriptaddress]{revtex4}
\usepackage{dcolumn,color,helvet,times}
\usepackage{amsmath,lastpage,bm,textcomp}
\usepackage{graphicx}
\usepackage{amsmath}
\usepackage{epsfig}
\usepackage{figcaps}

\begin{document}

\figcapsoff
\setcounter{page}{1} 
\title{Quantum limited particle sensing in optical tweezers}

\author{Jian~Wei~Tay}
\affiliation{Jack Dodd Centre for Photonics and Ultracold Atoms, Department of Physics, University of Otago, Dunedin, New Zealand}
\author{Magnus~T.~L.~Hsu}
\affiliation{School of  Mathematics and Physics, University of Queensland, St Lucia, QLD 4072, Australia}
\author{Warwick~P.~Bowen}
\affiliation{Jack Dodd Centre for Photonics and Ultracold Atoms, Department of Physics, University of Otago, Dunedin, New Zealand}
\affiliation{School of Mathematics and Physics, University of Queensland, St Lucia, QLD 4072, Australia}

\begin{abstract}
{Particle sensing in optical tweezers systems provides information on the position, velocity and force of the specimen particles. The conventional quadrant detection scheme is applied ubiquitously in optical tweezers experiments to quantify these parameters. In this paper we show that quadrant detection is non-optimal for particle sensing in optical tweezers and propose an alternative optimal particle sensing scheme based on spatial homodyne detection. A formalism for particle sensing in terms of transverse spatial modes is developed and numerical simulations of the efficacy of both quadrant and spatial homodyne detection are shown. We demonstrate that an order of magnitude improvement in particle sensing sensitivity can be achieved using spatial homodyne over quadrant detection.}{}{}
\end{abstract}
\maketitle

\section{INTRODUCTION}

The application of the radiation pressure force for the trapping of atoms and neutral particles was pioneered by Arthur Ashkin \cite{ashkin70s}. This was followed by a plethora of seminal experiments utilizing the radiation pressure force \cite{ashkin80}, for example in the displacement and levitation in air and water of micron-sized particles \cite{ashkin-levitate}, and together with Steve Chu, for the development of a stable three-dimensional atom cooling and trapping experiment using frequency-detuned counter-propagating laser beams \cite{chu-MOT}. In particular, the demonstration of {\it optical tweezers} \cite{ashkin-tweezer}, based largely on the transverse gradient force of a single focused Gaussian optical beam was a significant contribution to optical trapping in biology \cite{ashkin-tweezera}.

In biological systems, optical tweezers were first used to trap and manipulate viruses and bacteria \cite{bacteria}. This was followed by a burgeoning number of experiments using optical tweezers for measurements of DNA/RNA stretching and unfolding \cite{gore, bustamante, bustamantea, bryant-dna, smith}, intracellular probing, manipulation of gamete cells, trapping of vesicles, membranes and colloids \cite{langblock, neuman} and DNA sequencing using RNA polymerase \cite{greenleaf}. In particular, for the first time, quantitative biophysical studies of the kinetics of molecular motors \cite{bustamante-motor} (e.g. myosin \cite{myosin} and kinesin \cite{kinesin}) at the single molecule level was made possible with the use of optical tweezers. Coupled with conventional position sensitive detectors (i.e. using quadrant photodetectors \cite{langblock, gittes, pralle}), the position of, and force on, a bead tethered to a molecular motor can be measured at the single molecule level \cite{mehta-review, wang, moffitt}. The sensitivities attainable for force and position measurements of particles in optical tweezers are in the sub-piconewton and sub-nanometer regimes \cite{langblock, myosin, moffitt}, respectively. The application of the optical tweezers technology has led to a more complete biophysical understanding of the kinetics of molecular motors \cite{spudich, mehta, rosenfeld, spudich-cell, valentine, schnitzer, spudich-myosin6, bryant, dunn, okten, purcell, spudichrock} - a quintessential demonstration of new physical techniques yielding new insights into biology.

Beam position and momentum sensing is particularly crucial for particle sensing in optical tweezers enabling high-precision particle position and force measurements \cite{langblock, gittes, pralle}. Therefore it is important that such measurements are performed optimally to achieve the highest measurement efficacy. Recently, Hsu {\it et al.} \cite{HsuJOB} showed that the conventional quadrant detection scheme is non-optimal for measurements of the position and momentum of optical beams, even in the absence of classical noise sources. An alternative scheme for the optimal detection of the position and momentum of an optical beam was proposed, based on a spatial homodyne detection scheme. This scheme has also been proven to perform at the quantum limit of light based on Cramer-Rao informational bounds \cite{cramer-rao}. Therefore, it has become apparent that the use of quadrant detection for particle sensing in optical tweezers systems is non-optimal; and the introduction of spatial homodyne detection could offer the possibility for greater particle tracking sensitivities.

In this paper, we address the pertinent questions for particle sensing technology in optical tweezers systems - have we reached the limit of particle tracking sensitivity and can this limit be surpassed using quantum resources? We believe that in answering this technique related question, naturally arises a biophysical question - i.e. with significantly enhanced sensitivities, are we able to detect molecular kinetics, at the single molecule level, that were previously unresolvable? This biophysical question has wide implications as there are many vital protein conformational changes that occur in the angstrom regime, and within millisecond timescales \cite{moffitt}. For example, molecular motors move along nucleic acids in steps of a single-base pair scale (e.g. 3.4~\AA~on dsDNA) \cite{moffitt} and the bacterial DNA translocase FtsK moves at speeds of 5 kilobases per second \cite{pease}. Therefore, enhanced particle sensing could elucidate these finer features with greater sensitivity than conventional particle sensing techniques in optical tweezers systems.

This paper begins by formalizing an optimal parameter estimation procedure for particle sensing based on the analysis of the spatial properties of the field scattered by a particle in an optical tweezers. We show that split detection is non-optimal and consequently propose an optimal measurement scheme based on spatial homodyne detection. The efficacy of particle sensing is evaluated using the signal-to-noise ratio (SNR) and sensitivity measures; and the efficacy of spatial homodyne detection and split detection systems are compared.

\section{OPTIMAL PARAMETER ESTIMATION FOR SPATIAL PROPERTIES OF OPTICAL FIELDS\label{sec:estimate}}

An optical field can be formalized and described using a range of parameters - e.g. the polarization, the amplitude-phase quadratures, and the transverse spatial profile. These parameters can be measured using a range of detection techniques (e.g. polarimetry, direct detection, interferometry and beam profiling) and an estimate of their values in the presence of classical and quantum noise, and detection inefficiency is obtained. Here we develop a formalism to quantify an arbitrary spatial modification of the field parameterized by a parameter $p$ (e.g. $p$ could quantify the displacement of a spatial mode along a transverse axis \cite{HsuJOB}). In principle, an arbitrary field can be treated and the field properties can be modeled using Maxwell's equations \cite{mie}. However, for spherical fields such as those produced by scattering processes from small particles, after optical imaging of the field, the paraxial approximation is valid and the propagating field can be described using two-dimensional spatial modes in a convenient basis.

The sensitivity of measurements on optical fields is ultimately limited by quantum noise on the fields, exhibited typically as shot noise. To understand such limits it is important to use a full quantum mechanical description of the field. The spatial quantum states of an optical field exist within an infinite dimensional Hilbert space. Depending on the spatial symmetry of an imaged optical field, the spatial states of the field may be conveniently expanded in the basis of the rectangularly-symmetric TEM$_{mn}$ or circularly-symmetric LG$_{n,l}$ modes. A field of frequency $\omega$ can be represented by the positive frequency part of the electric field operator $\tilde{\bf E}^{+}({\mbox{\boldmath$\rho$}})e^{i\omega t}$. We are interested in the transverse information of the field described fully by the slowly varying field envelope operator $\tilde{\bf E}^{+}({\mbox{\boldmath$\rho$}})$, given by
\begin{equation}  \label{spatialfield1}
\tilde{\bf E}^{+} ({\mbox{\boldmath$\rho$}}) = i \sqrt{\frac{\hbar \omega}{2\epsilon_{0} V}} \sum_{j,m,n} \tilde{a}_{mn}^{j} {\bf u}_{mn}^{j} ({\mbox{\boldmath$\rho$}}),
\end{equation}
where ${\mbox{\boldmath$\rho$}} = (x,y)$ is a co-ordinate in the transverse plane of the field, and the summation over the parameters $j$, $m$ and $n$ is given by
\begin{equation}
\sum_{j,m,n} \equiv \sum_{j \in \{x,y\}} \sum_{m=0}^{\infty} \sum_{n=0}^{\infty}.
\end{equation}
In this paper, we adopt the TEM$_{mn}$ mode basis for convenience, such that $\mathbf{u}_{mn}^j ({\mbox{\boldmath$\rho$}})$ and $\tilde{a}_{mn}^j$ are, respectively, the transverse beam amplitude function and the photon annihilation operator for the TEM$_{mn}$ mode with polarization $j$. The ${\bf u}_{mn} ({\mbox{\boldmath$\rho$}})$ mode functions are normalized such that their self-overlap integrals are unity, so that the inner product
\begin{eqnarray}
\left \langle {\bf u}^j_{mn} ({\mbox{\boldmath$\rho$}}), {\bf u}^{j'}_{m'n'} ({\mbox{\boldmath$\rho$}}) \right \rangle & = &  \int_{-\infty}^{\infty} \Big [ {\bf u}^j_{mn} ({\mbox{\boldmath$\rho$}}) \Big ]^{*} \cdot {\bf u}^{j'}_{m'n'} ({\mbox{\boldmath$\rho$}}) d {\mbox{\boldmath$\rho$}} \nonumber\\
& = &\delta_{mm'} \delta_{nn'} \delta_{jj'}.
\end{eqnarray}

We now apply an arbitrary spatial perturbation, described by parameter $p$, to the field. Eq.~(\ref{spatialfield1}) can then be rewritten as a sum of coherent amplitude components and quantum noise operators, given by
\begin{eqnarray}  \label{spatialfield}
\tilde{\bf E}^{+} ({\mbox{\boldmath$\rho$}},p) & = &i \sqrt{\frac{\hbar \omega}{2\epsilon_{0} V}} \sum_{j,m,n} \tilde{a}_{mn}^{j} {\bf u}_{mn}^{j} ({\mbox{\boldmath$\rho$}},p) \nonumber\\
& = & i \sqrt{\frac{\hbar \omega}{2\epsilon_{0} V}} \Big[ \alpha (p) {\bf v}({\mbox{\boldmath$\rho$}},p) + \sum_{j,m,n} \delta \tilde{a}_{mn}^j {\bf u}_{mn}^j ({\mbox{\boldmath$\rho$}},0) \Big],  \nonumber\\
\end{eqnarray}
where
$$\alpha (p) {\bf v}({\mbox{\boldmath$\rho$}},p) = \sum_{j,m,n} \langle \tilde{a}_{mn}^j \rangle {\bf u}_{mn}^j ({\mbox{\boldmath$\rho$}},0) = \sum_{j,m,n} \langle \tilde{a}_{mn}^j \rangle u_{mn} ({\mbox{\boldmath$\rho$}},0) \hat{\bf j},$$
$\alpha (p) $ is the coherent amplitude of mode ${\bf v}({\mbox{\boldmath$\rho$}},p)$, and $\hat{\bf j}$ is the unit polarization vector. We see from Eq.~(\ref{spatialfield}) that $\alpha (p)$ and ${\bf v}({\mbox{\boldmath$\rho$}},p)$ can be related to $\tilde{\bf E}^{+} ({\mbox{\boldmath$\rho$}},p)$ by
\begin{eqnarray}
\alpha(p) & = &\sqrt{\frac{2 \epsilon_{0} V}{\hbar \omega} \left \langle \overline{\bf E}^{+}({\mbox{\boldmath$\rho$}},p),\overline{\bf E}^{+}({\mbox{\boldmath$\rho$}},p) \right \rangle } \label{alphap} \\
{\bf v}({\mbox{\boldmath$\rho$}},p) & = & - i N_v \overline{\bf E}^{+}({\mbox{\boldmath$\rho$}},p)\label{v}
\end{eqnarray}
where $\overline{\bf E}^{+}({\mbox{\boldmath$\rho$}},p) = \langle \tilde{\bf E}^{+}({\mbox{\boldmath$\rho$}},p) \rangle$, and the normalization constant $N_v $ is given by
\begin{eqnarray} \label{ortho}
N_v & = & \left \langle \overline{\bf E}^{+}({\mbox{\boldmath$\rho$}},p),\overline{\bf E}^{+}({\mbox{\boldmath$\rho$}},p) \right \rangle^{-1/2} \nonumber\\
& = & \left[ \iint_{-\infty}^{\infty} \Big [ \overline{\bf E}^{+}({\mbox{\boldmath$\rho$}},p) \Big]^{*} \cdot \overline{\bf E}^{+}({\mbox{\boldmath$\rho$}},p) d {\mbox{\boldmath$\rho$}} \right]^{-1/2} .
\end{eqnarray}
Note that $|\alpha (p) |^2$ is the mean number of photons passing through the transverse plane of the field per second and in this paper we assume $\alpha(p)$ to be real, without loss of generality. The quantum noise operator corresponding to mode ${\bf u}_{mn} ({\mbox{\boldmath$\rho$}}) = {\bf u}^j_{mn}({\mbox{\boldmath$\rho$}},0)$ is given by $\delta \tilde{a}_{mn}^j$.

In the limit of small estimate parameter $p$, the Taylor expansion of the first bracketed term in Eq.~(\ref{spatialfield}) is given by
\begin{equation} \label{taylor}
\alpha (p) {\bf v}({\mbox{\boldmath$\rho$}},p) \approx \alpha (0) {\bf v}({\mbox{\boldmath$\rho$}},0) + p \cdot \frac{\partial [\alpha (p) {\bf v}({\mbox{\boldmath$\rho$}},p)] }{\partial p} \bigg |_{p=0} ,
\end{equation}
where the first term on the right-hand side of Eq.~(\ref{taylor}) indicates that the majority of the power of the field is in the ${\bf v}({\mbox{\boldmath$\rho$}},0)$ mode. The second term on the right-hand side of Eq.~(\ref{taylor}) defines the spatial mode ${\bf w}({\mbox{\boldmath$\rho$}})$ corresponding with small changes in the parameter $p$
\begin{equation} \label{w}
{\bf w}({\mbox{\boldmath$\rho$}})= \frac{\partial [\alpha (p) {\bf v}({\mbox{\boldmath$\rho$}},p)] }{\partial p} \bigg |_{p=0}.
\end{equation}
From Eq.~(\ref{taylor}) we see that the amplitude of mode ${\bf w}({\mbox{\boldmath$\rho$}})$ is directly proportional to the magnitude of the spatial perturbation of the field.

\subsection{Split detection}

\begin{figure}[!ht]
\begin{center}
\includegraphics[width=7cm,keepaspectratio]{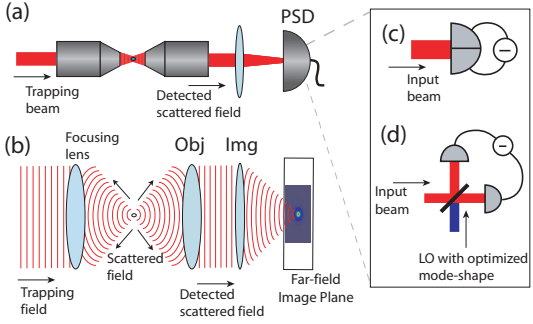}
\caption{(a) Schematic diagram and (b) wavefront illustration for an optical tweezers experiment. A trapping field is focused onto a particle. The particle scatters the incident trapping field, with the resulting scattered and residual trapping fields collected by an objective (Obj) lens. This is followed by imaging of the collected fields onto a position sensitive detector (PSD) in the far-field using an imaging (Img) lens. The position sensitive detector could consist of either a (c) split detection system or a (d) spatial homodyne scheme with the local oscillator (LO) beam in an optimized spatial mode for the relevant measurement of parameter $p$.}
\label{schematic}
\end{center}
\end{figure}

For optical beam position and momentum measurements, the conventional detection scheme used is split detection (a one-dimensional quadrant detector). In split detection, the optical beam under interrogation is incident centrally on a split detector, as shown in Fig.~\ref{schematic}~(c).
The difference between the photocurrents from the two halves of the split detector contains partial information about the position/momentum of the beam, given by \cite{HsuJOB}
\begin{eqnarray}
\Delta i_{\rm SD} & = & \frac{2\epsilon_{0} V}{\hbar \omega} \left[ \int_{-\infty}^{0} \tilde{{\bf E}}^{+ \dagger} \cdot \tilde{{\bf E}}^{+} d {{\mbox{\boldmath$\rho$}}} - \int_{0}^{\infty} \tilde{{\bf E}}^{+ \dagger} \cdot \tilde{{\bf E}}^{+} d {{\mbox{\boldmath$\rho$}}} \right] \nonumber\\
&=& \alpha(p) \tilde X^+_f ,
\end{eqnarray}
where $\tilde X^+_f = \tilde a_f^\dagger + \tilde a_f$ is the amplitude quadrature operator of the flipped mode with transverse mode amplitude function
\begin{equation}
{\bf v}_f({{\mbox{\boldmath$\rho$}}}) = \left\{ {\begin{array}{*{20}c}
   {{\bf v}({{\mbox{\boldmath$\rho$}}},0),} & {x \ge 0}  \\
   { - {\bf v}({{\mbox{\boldmath$\rho$}}},0),} & {x < 0}  \\
\end{array}} \right.
\end{equation}
The amplitude quadrature operator can be written in terms of its coherent amplitude
\begin{equation}
\alpha_f (p) = \alpha(p) \left \langle {\bf v}_f({{\mbox{\boldmath$\rho$}}}), {\bf v}({{\mbox{\boldmath$\rho$}}}, p) \right \rangle \label{overlapSD}
\end{equation}
wherein resides the signal due to the parameter $p$, and a quantum noise operator $\delta \tilde X^+_f = \tilde X^+_f - \langle \tilde X^+_f \rangle$ which is ultimately responsible for placing a quantum limit on the measurement sensitivity so that
\begin{equation}
\Delta i_{\rm SD}  = \alpha(p) \left [ 2 \alpha_f (p) + \delta \tilde X^+_f \right ].
\end{equation}
%
%
%
The $\langle {\bf v}_f({{\mbox{\boldmath$\rho$}}}), {\bf v}({{\mbox{\boldmath$\rho$}}}, p) \rangle$ term in Eq.~(\ref{overlapSD}) is the overlap integral between the flipped mode ${\bf v}_f({{\mbox{\boldmath$\rho$}}})$ and the displaced mode ${{\bf v}({{\mbox{\boldmath$\rho$}}},p)}$.

\subsection{Spatial homodyne detection}

Hsu {\it et al.} \cite{HsuJOB} proposed a new displacement measurement scheme that is optimal for detecting beam position and momentum. The spatial homodyne scheme utilizes a homodyne detection setup that has a local oscillator mode optimized for the displacement measurement of the input beam, as shown in Fig.~\ref{schematic}~(d). The local oscillator (LO) beam interferes with the input beam on a 50/50 beam-splitter. The outputs of the beam-splitter are then detected using a pair of balanced single-element photodetectors, with the difference in photocurrents providing the measurement signal. The spatial homodyne scheme was also proven to perform at the Cramer-Rao bound \cite{cramer-rao}, therefore extending the capabilities of the spatial homodyne scheme for the optimal measurement of any spatial parameter $p$ (e.g. the measurement of the orbital angular momentum of light \cite{hsu-stokes}). We now proceed to derive the photocurrent for the spatial homodyne detection scheme.

The input beam (as described in Eq.~(\ref{spatialfield})) is interfered with the bright LO beam with mode-shape ${\bf w}({{\mbox{\boldmath$\rho$}}})$. The positive frequency part of the electric field operator for the LO given by
\begin{eqnarray} \label{LOspatialfield}
\tilde{{\bf E}}^{+}_{\rm LO} ({\mbox{\boldmath$\rho$}}) & = & i \sqrt{\frac{\hbar \omega}{\epsilon_{0} c}} \Big [ \alpha_{\rm LO} {\bf w} ({\mbox{\boldmath$\rho$}}) + \sum_{j,m,n} \delta \tilde{a}_{mn,{\rm LO}}^j {\bf u}_{mn}^j ({\mbox{\boldmath$\rho$}}) \Big ] e^{i \phi},\nonumber\\
\end{eqnarray}
where $\phi$ is the phase difference between the local oscillator and the input beam.

The photocurrent at each photodetector (distinguished by the subscripts + and -, respectively), assuming detectors of infinite extent, is given by
\begin{eqnarray}
i_{\pm} &=& \frac{2\epsilon_{0} V}{\hbar \omega} \int_{-\infty}^{\infty} \tilde{{\bf E}}^{+ \dagger}_{\pm} \cdot  \tilde{{\bf E}}^{+}_{\pm} d {{\mbox{\boldmath$\rho$}}}\\
&=& \frac{2\epsilon_{0} V}{\hbar \omega} \int_{-\infty}^{\infty} (\tilde{{\bf E}}^{+} \pm \tilde{{\bf E}}^{+}_{\rm LO})^\dagger  \cdot (\tilde{{\bf E}}^{+} \pm \tilde{{\bf E}}^{+}_{\rm LO}) d {{\mbox{\boldmath$\rho$}}},\nonumber\\ \label{temp1}
\end{eqnarray}
whereby one output of the spatial homodyne attains a $\pi$ phase shift with respect to the other output due to the hard-reflection from the beam-splitter.

Substituting Eqs.~(\ref{spatialfield1}) and (\ref{LOspatialfield}) into Eq.~(\ref{temp1}) and taking the subtraction of the photocurrent from the two detectors gives
\begin{eqnarray} \label{photohomo}
\Delta i_{\rm SH} &=& i_{+}-i_{-} \nonumber \\
& = & \alpha_{\rm LO}  \int_{-\infty}^{\infty}  \Big[ e^{-i \phi} [ {\bf w} ({{\mbox{\boldmath$\rho$}}}) ]^* \cdot \sum_{j,m,n} \tilde{a}_{mn}^j {\bf u}_{mn}^j ({{\mbox{\boldmath$\rho$}}},p) \nonumber\\
& & + e^{i \phi} {\bf w} ({{\mbox{\boldmath$\rho$}}}) \cdot \left ( \sum_{j,m,n}  \tilde{a}_{mn}^{j} {\bf u}_{mn}^j ({{\mbox{\boldmath$\rho$}}},p) \right )^\dagger \Big] d {{\mbox{\boldmath$\rho$}}} \nonumber \\
&=& \alpha_{\rm LO}  \Big[  e^{-i \phi} \sum_{j,m,n} \tilde{a}_{mn}^j \left \langle {\bf w} ({{\mbox{\boldmath$\rho$}}}), {\bf u}_{mn}^j ({{\mbox{\boldmath$\rho$}}},p) \right \rangle \nonumber\\
& & + e^{i \phi}  \left( \sum_{j,m,n} \tilde{a}_{mn}^{j} \left \langle {\bf w} ({{\mbox{\boldmath$\rho$}}}), {\bf u}_{mn}^j ({{\mbox{\boldmath$\rho$}}},p) \right \rangle \right )^\dagger \Big] \nonumber \\
&=& \alpha_{\rm LO} \left[  e^{-i \phi} \tilde{a}_w  + e^{i \phi}  \tilde{a}_w^\dagger  \right] \nonumber \\
&=& \alpha_{\rm LO} \tilde X^{\phi}_{w}
\end{eqnarray}
where $\tilde a_w$ is an annihilation operator describing the component of the input field in mode ${\bf w} ({{\mbox{\boldmath$\rho$}}})$, and by definition the $\tilde X^{\phi}_{w}=e^{-i \phi} \tilde{a}_w  + e^{i \phi}  \tilde{a}_w^\dagger $ is the quadrature operator of that component at phase angle $\phi$. In the above, we have taken the condition $\alpha_{\rm LO} \gg \langle \tilde{a}_w \rangle$ and invoked the linearization approximation, thereby removing terms that do not involve $\alpha_{\rm LO}$. The orthonormality property of modes given in Eq.~(\ref{ortho}) has also been used.

An optimal estimate of the parameter $p$ is obtained when the local oscillator mode ${\bf w}({{\mbox{\boldmath$\rho$}}})$ is chosen to match the associated input mode ${\bf v}({{\mbox{\boldmath$\rho$}}},p)$, as shown in Eq.~(\ref{photohomo}). The spatial homodyne detection scheme then extracts from the signal field a quadrature variable associated with the local oscillator field mode, with quadrature phase angle given by $\phi$.

It should be noted that Delaubert {\it et al.} \cite{cramer-rao} have shown that optimal parameter estimation can be achieved using a photodetector array for the cases where the signal field $\tilde{\bf E}^{+} ({\mbox{\boldmath$\rho$}},p)$ is shot noise limited or single mode squeezed, so long as the array resolution is sufficiently small. Array detection is restricted to amplitude quadrature detection and is not polarization resolving, however in situations where these restrictions are satisfied is formally identical to spatial homodyne detection.


\subsection{Quantifying the efficacy of parameter estimation}

We now introduce the SNR and sensitivity measures for the spatial homodyne and split detection schemes.

For the spatial homodyne detection scheme, the measured signal is the mean signal component of the difference photocurrent in Eq.~(\ref{photohomo}), given by
\begin{equation}
\langle \Delta i_{\rm SH} \rangle = \alpha_{\rm LO} \alpha_w (p) \left ( e^{i \Delta \phi} + e^{-i \Delta \phi} \right ),
\end{equation}
where $\alpha_w(p) = \alpha(p) \left \langle {\bf w}({{\mbox{\boldmath$\rho$}}}), {\bf v}({{\mbox{\boldmath$\rho$}}}, p)\right \rangle$. For matched local oscillator and signal phases such that $\phi = 0$, the maximal signal is obtained, given by
\begin{equation}
\langle \Delta i_{\rm SH} \rangle = 2 \alpha_{\rm LO} \alpha_w(p).
\end{equation}
The corresponding noise component is given by
\begin{equation}
\sqrt{ \langle \Delta i_{\rm SH}^2 \rangle - \langle \Delta i_{\rm SH} \rangle^2} = \alpha_{\rm LO} \Delta \tilde X^{\phi}_{w},
\end{equation}
where $\Delta^2 \tilde X^{\phi}_{w} = \langle ( \delta \tilde X^{\phi}_{w} )^2 \rangle$ is the variance of the signal field mode. The resulting SNR is given by
\begin{equation}
{\rm SNR_{\rm SH}} = \frac{2 \alpha_w (p)}{\Delta \tilde X^{\phi_{\rm LO}}_{w}}.
\end{equation}
If the optical field is in a coherent state, as is typical of a low noise laser $\Delta \tilde X^{\phi_{\rm LO}}_{w}=1$ and the SNR for the spatial homodyne detection scheme is given by
\begin{equation} \label{optimalSNR}
{\rm SNR_{\rm SH,coh}} =  2 \alpha_w(p).
\end{equation}
Clearly, although experimentally challenging, squeezing the signal mode such that $\Delta \tilde X^{\phi_{\rm LO}}_{w}<1$ has the capacity to further enhance the SNR.

Alternatively, we introduce the sensitivity $\mathcal{S}$ measure, which is defined as a change to parameter $p$ required to provide ${\rm SNR} =1$ for a signal field in a coherent state, given by
\begin{equation} \label{sensSH}
\mathcal{S}_{\rm SH,coh} = \left [ \frac{\partial {\rm SNR}}{\partial p} \bigg |_{p=0} \right ]^{-1} = \frac{1}{2} \left [ \frac{\partial \alpha_w (p)}{\partial p} \bigg |_{p=0} \right ]^{-1}.
\end{equation}

For comparison, the corresponding SNR for the split detection scheme in the coherent state limit is given by
\begin{equation} \label{snrSD}
{\rm SNR}_{\rm SD, coh} = 2 \alpha_f(p),
\end{equation}
with a sensitivity given by
\begin{equation} \label{sensSD}
\mathcal{S}_{\rm SD,coh} =  \frac{1}{2} \left [ \frac{\partial \alpha_f (p)}{\partial p} \bigg |_{p=0} \right ]^{-1}.
\end{equation}
%

\section{PARTICLE SENSING IN OPTICAL TWEEZERS\label{sec:opttweezer}}

Fig.~\ref{schematic} (a) shows a typical optical tweezers setup. A trapping beam in the TEM$_{00}$ mode is focused onto a scattering particle. In this instance we assume that the particle is spherical, with a permittivity greater than that of the medium, $\epsilon_2 > \epsilon_1$. If the particle has a diameter larger than the wavelength of the trapping beam, light rays are refracted as they pass through the particle, as shown in Fig. 2. This refracted light results in an equal and opposite change of momentum imparted on the particle. Due to the intensity profile of the beam, the outer ray is less intense than the inner ray. Consequently, the resulting force acts to return the particle to the center of the trapping beam focus \cite{ashkin-tweezera}.
\begin{figure}[!ht]
\begin{center}
\includegraphics[width=6.5cm]{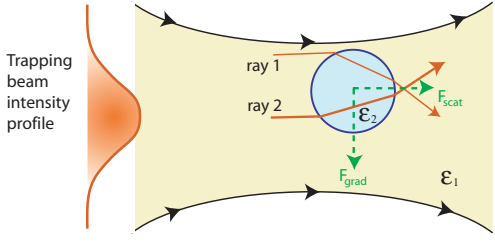}
\caption{Illustration showing a TEM$_{00}$ trapping beam impinging on a spherical scattering particle. Rays 1 and 2 are refracted in the spherical particle, thereby undergoing a change in momentum. A corresponding equal and opposite change in momentum is imparted on the particle resulting in the particle being attracted to the center of the trapping beam. ${\bf F}_{\rm grad}$ and ${\bf F}_{\rm scat}$ are the gradient and scattering forces, respectively. $\epsilon_1$ and $\epsilon_2$ are the respective permittivity of the medium and the sample.}
\label{tweezer}
\end{center}
\end{figure}

The effective restoring/trapping force is due to two force components - (i) the {\it gradient force} ${\bf F}_{\rm grad}$ resulting from the intensity gradient of the TEM$_{00}$ trapping beam, that acts transversely toward the high intensity region and (ii) the {\it scattering force} ${\bf F}_{\rm scat}$ resulting from the forward-direction radiation pressure of the trapping beam incident on the particle. In the focal region of the optical tweezers trap the gradient force is typically dominant.

It is important to note that in some optical tweezers experiments the trapped particle has radius less than the wavelength of the trapping laser. In this regime, the trapping force on the particle is generated due to an induced dipole moment. The dipole moment induced will be along the direction of trapping beam polarization. The assumption that the particle is spherical is no longer important, since the particle has no structural deviations greater than the wavelength of the trapping beam. This allows the particle to be treated as a normal dipole, hence the particle experiences a force due to interaction of its induced dipole moment with the transverse electromagnetic fields of the impinging light. This force is proportional to the intensity of the beam and has the same net result as before; it acts to return the particle to the center of the trapping beam focus.

The position and force sensing of the trapped particle can then be obtained by imaging the scattered field from the particle on a position sensitive detector such as the commonly utilized quadrant photo-detector \cite{langblock, gittes, pralle}, or a spatial homodyne detector\cite{HsuJOB}.

\subsection{System configuration}

The collection efficiency of the light field is given by the numerical aperture (NA) of the objective lens (as shown in Fig.~\ref{schematic}~(b)), given by
\begin{equation}
{\rm NA} = n \sin\theta,
\end{equation}
where $n$ and $\theta$ are the refractive index and the collection half-angle of the lens, respectively. $\theta$ is related to the lens diameter $D$ (assuming the object is at the focus, with focal length $f_L$) by
\begin{equation}
\tan\theta = \frac{D/2}{f_L}.
\end{equation}

\subsection{Propagation of fields through system}

We now formalize all the relevant fields that propagate through the optical tweezers system, as shown in the schematic of the optical tweezers arrangement of Fig.~\ref{schematic}~(a). Fig.~\ref{schematic}~(b) illustrates the wave-front of the trapping and scattered fields. The trapping field is incident from the left of the diagram and is then focused onto a spot, from the focusing lens. The particle is trapped near the center of this focal spot and scatters the incident trapping field, with the forward scattered and residual trapping field being collected by the objective lens. This is followed by imaging into the far-field onto a position sensitive detector.

\subsubsection*{Trapping field}

Assuming that the trapping field is Gaussian and hence in a TEM$_{00}$ mode, the positive frequency part of the electric field for the trapping beam at the waist of the trap (denoted by the superscript ${\rm T}$), is given by
\begin{equation} \label{eqn:GaussianBeam}
\overline{\bf E}^{{\rm T}+}_{\rm trap}({\mbox{\boldmath$\rho$}}) = i \sqrt{\frac{\hbar \omega}{2\epsilon_{0} V}} \alpha_{\rm trap} {\bf u}^{\rm T}_{00}({\mbox{\boldmath$\rho$}}),
\end{equation}
with the mode-shape function given by
\begin{equation}
{\bf u}^{\rm T}_{00}({\mbox{\boldmath$\rho$}}) = \frac{2}{w_{\rm T} \sqrt{\pi}} e^{-{\rho^2}/{w_{\rm T}^2}} \hat {\bf p}_{\rm trap},
\end{equation}
where $\rho^2 = | {\mbox{\boldmath$\rho$}}|^2$, $w_{\rm T}$ is the waist size of the trapping beam, and $\hat {\bf p}_{\rm trap}$ is a unit vector representing the polarization of the trapping field.

Using the paraxial approximation, the positive frequency part of the electric field of the trapping beam after propagation of a distance $f_{\rm O}$ from the focus to the objective lens (of focal length $f_{\rm O}$) is given by
\begin{equation}
\overline{\bf E}^{{\rm O} +}_{\rm trap}({\mbox{\boldmath$\rho$}}) =  i \sqrt{\frac{\hbar \omega}{2\epsilon_{0} V}} \alpha_{\rm trap} e^{-i k f_{\rm O}} {\bf u}^{\rm O}_{00} \Pi_{R}({\mbox{\boldmath$\rho$}}),
\end{equation}
where $k=2 \pi/\lambda$ is the wave-vector of the trapping field. With the exception of the replacement $w_{\rm O} \rightarrow w_{\rm T}$, ${\bf u}^{\rm O}_{00} ({\mbox{\boldmath$\rho$}})$ is defined identically to ${\bf u}^{\rm T}_{00}$, with the radius of the spot at the objective being $w_{\rm O}$ given by
\begin{equation} \label{eqn:FocalSpot}
w_{\rm O} = \frac{f_{\rm O} \lambda}{\pi w_{\rm T}}.
\end{equation}
Aperturing due to the finite radius $R$ of the objective lens is taken into account via the aperture function $\Pi_R({\mbox{\boldmath$\rho$}})$ given by
\begin{equation}
\Pi_R ({\mbox{\boldmath$\rho$}}) = \bigg{\{} {\begin{array}{*{20}c}
   {1,} & {\rho < R}  \\
   { 0,} & {\rho \ge 0} ,\\
\end{array}}
\end{equation}
where $R$ can be related to the numerical aperture (NA) of the imaging system and refractive index of the trapping medium $n$ by $R= f_{\rm O} {\rm NA} / \sqrt{n^2 - {\rm NA}^2}$.

\subsubsection*{Scattered field}

In principle, there could be multiple inhomogeneous particles within the optical tweezers focus, scattering the input trapping field. For this scenario, several numerical methods exist to calculate the scattered field - e.g. the finite difference frequency domain and T-matrix hybrid method \cite{loke07} and the discrete-dipole approximation and point matching method \cite{loke09}. However, for simplicity we consider the scattering from a single spherical, homogeneous particle with diameter much smaller than the wavelength. The resulting scattered field can be modeled as dipole radiation, having a positive frequency electric field  \cite{mie} given by
\begin{equation} \label{eqn:ScattField}
\overline{\bf E}_{\rm scat}^+({\bf r},p) = - k^2 a^3  \left( \frac{\epsilon_1 - \epsilon_2}{\epsilon_1 + 2 \epsilon_2} \right) \frac{e^{-ik r'}}{r'} \hat {\bf r}' \times
\hat {\bf r}' \times \overline{\bf E}_{\rm trap}^{{\rm T} +}(p),
\end{equation}
where $\mathbf{r} = x \hat {\bf x} + y \hat {\bf y} + z \hat {\bf z}$ is the coordinate of the field with respect to the center of the optical tweezers, $\mathbf{r}' = (x-p) \hat {\bf x} + y \hat {\bf y} + z \hat {\bf z}$ is the coordinate of the field with respect to the displaced particle, $r'=|{\bf r}'|$, and $\hat {\bf r}'={\bf r}'/r'$. The radius of the spherical scattering particle is given by $a$.

The scattered field is then collected by the objective lens (as shown in Fig.~\ref{schematic}~(b)), with the corresponding positive frequency part of the electric field given by
\begin{eqnarray} \label{eqn:ScattFieldObj}
\overline{\bf E}_{\rm scat}^{{\rm O}+} ({\mbox{\boldmath$\rho$}},p) &=& \left [ \left  ( \overline{\bf E}_{\rm scat}^+({\bf r}_{\rm O},p) \cdot \hat {\bf l} \right )\hat {\bf l} + \left (\overline{\bf E}_{\rm scat}^+({\bf r}_{\rm O}) \cdot \hat {\bf m} \right ) \hat {\bf n} \right ] \nonumber\\
& & \cdot \sqrt{\frac{f_{\rm O}}{r_{\rm O}'}} e^{i k (r_{\rm O} - f_{\rm O})} \Pi_{R}({\mbox{\boldmath$\rho$}})\\
&=& -i K \sqrt{\frac{\hbar \omega}{\epsilon_{0} c}} \sqrt{\frac{f_{\rm O}}{r_{\rm O}'}} \frac{e^{-ik (r_{\rm O}-r_{\rm O}'-f_{\rm O})}}{r_{\rm O}'} \Pi_{R}({\mbox{\boldmath$\rho$}}) \nonumber\\
& & \cdot \Big[ \left  ( \hat {\bf r_{\rm O}}' \times \hat {\bf r_{\rm O}}' \times {\bf u}^{\rm O}_{00} \right )\hat {\bf l} + \left (\hat {\bf r_{\rm O}}' \times \hat {\bf r_{\rm O}}' \times {\bf u}^{\rm O}_{00} \right ) \hat {\bf n} \Big], \nonumber\\
\end{eqnarray}
where ${\bf r}_{\rm O} = x \hat {\bf x} + y \hat {\bf y} + f_{\rm O} \hat {\bf z}$ and ${\bf r}_{\rm O}' = (x-p) \hat {\bf x} + y \hat {\bf y} + f_{\rm O} \hat {\bf z}$. The unit vectors $\mathbf{\hat{l}}$, $\mathbf{\hat{m}}$, and $\mathbf{\hat{n}}$
\begin{eqnarray}
\mathbf{\hat{l}} &=& \frac{1}{\rho'}(y, p - x, 0) \\
\mathbf{\hat{m}} &=& \frac{1}{\rho' r_{\mathrm O}}(-f_{\rm O}(x-p), - f_{\rm O} y,\rho'^2)\\
\mathbf{\hat{n}} &=& \frac{-1}{\rho'}(x-p,y,0),
\end{eqnarray}
are used to include the effect of the objective lens on the polarization of the scattered field, where
${\mbox{\boldmath$\rho$}}' = (x - p) \hat {\bf x} + y \hat {\bf y}$ and $\rho' = |{\mbox{\boldmath$\rho$}}'|$. The term $\sqrt{{f_{\rm O}}/{r_{\rm O}'}}$ describes the compression of the intensity of the scattered field due to the change in propagation direction induced by the objective lens. To simplify the equation, we have defined the constant $K$ given by
\begin{equation}
K= \alpha_{\rm trap} k^2 a^3 \left( \frac{\epsilon_1 - \epsilon_2}{\epsilon_1 + 2 \epsilon_2} \right).
\end{equation}

\subsubsection*{Detection}
Since the total field after the objective lens consists of both the scattered field and the residual trapping field, we now include both fields to describe the total field after the objective lens, given by
\begin{equation} \label{etot}
\overline{\bf E}_{\rm total}^{{\rm O}+} ({\mbox{\boldmath$\rho$}},p) = \overline{\bf E}_{\rm scat}^{{\rm O}+} ({\mbox{\boldmath$\rho$}},p) + \overline{\bf E}^{{\rm O} +}_{\rm trap}({\mbox{\boldmath$\rho$}}).
\end{equation}

After the objective lens, the beam is focused onto a detector in the far-field image plane, via the use of an imaging lens. Assuming the lens is thin and ideal, the field in the image plane is obtained by taking the Fourier transform of Eq.~(\ref{etot}), given by
\begin{eqnarray}
\overline{\bf E}_{\rm total}^{{\rm I} +} ({\mbox{\boldmath$\Gamma$}},p) &=& {\cal F} \left ( \overline{\bf E}^{{\rm O}+}_{\rm total}({\mbox{\boldmath$\rho$}},p)  \right )\\
&=& {\cal F} \left ( \overline{\bf E}_{\rm scat}^{{\rm O}+} ({\mbox{\boldmath$\rho$}} , p) \right ) + {\cal F} \left (\overline{\bf E}^{{\rm O} +}_{\rm trap}({\mbox{\boldmath$\rho$}}) \right )\\
&=& \overline{\bf E}_{\rm scat}^{{\rm I}+} ({\mbox{\boldmath$\Gamma$}},p) + \overline{\bf E}^{{\rm I} +}_{\rm trap}({\mbox{\boldmath$\Gamma$}}) \label{farfieldintensity}
\end{eqnarray}
where ${\mbox{\boldmath$\Gamma$}}=(X,Y)$ are the transverse co-ordinates in the image plane. It is important to note that the analysis presented here is independent of the absolute scaling of the image plane co-ordinates. In an experimental situation a scaling factor is introduced that depends on the choice of magnification lenses used.

The critical parameters for assessing sensitivity of particle monitoring are $\alpha(p)$, ${\bf v} ({\mbox{\boldmath$\Gamma$}},p)$ and ${\bf w} ({\mbox{\boldmath$\Gamma$}})$. These parameters can now be calculated using Eqs.~(\ref{alphap}), (\ref{v}) and (\ref{w}). Using Eq.~(\ref{alphap}) we now find
\begin{eqnarray}
\alpha(p) &=& \sqrt{\frac{2 \epsilon_{0} V}{\hbar \omega} \left \langle \overline{\bf E}_{\rm total}^{{\rm I} +}({\mbox{\boldmath$\Gamma$}},p),\overline{\bf E}_{\rm total}^{{\rm I} +}({\mbox{\boldmath$\Gamma$}},p) \right \rangle }\\
&\approx&  \sqrt{\frac{2\epsilon_{0} V}{\hbar \omega} \left \langle \overline{\bf E}_{\rm trap}^{{\rm I} +}({\mbox{\boldmath$\Gamma$}}),\overline{\bf E}_{\rm trap}^{{\rm I} +}({\mbox{\boldmath$\Gamma$}}) \right \rangle }\\
&=& \alpha_{\rm trap}, \label{atrap}
\end{eqnarray}
where we have assumed that the trap power is greater than the scattered power, as is the case for scattering from a small particle; and for simplicity that only the scattered field is apertured by the objective lens. The latter assumption is reasonable for optical tweezers systems with a sufficiently large trap waist size and numerical aperture. In this paper, we restrict our analysis to the realistic scenario of $\mathrm{NA}>0.2$ and choose a trapping field waist size of 4 $\mu$m.  With these parameters, trap field clipping due to the aperture causes only 15 ppm loss and is therefore negligible.

Using Eq.~(\ref{v}) we obtain
\begin{eqnarray}
{\bf v} ({\mbox{\boldmath$\Gamma$}},p) & = & -i N_v \overline{\bf E}_{\rm total}^{{\rm I} +} ({\mbox{\boldmath$\Gamma$}},p) \nonumber\\
& = & -i \sqrt{\frac{2\epsilon_0 V}{\hbar \omega}} \frac{1}{\alpha_{\rm trap}} \left ( \overline{\bf E}_{\rm scat}^{{\rm I}+} ({\mbox{\boldmath$\Gamma$}},p) + \overline{\bf E}^{{\rm I} +}_{\rm trap}({\mbox{\boldmath$\Gamma$}}) \right ),\nonumber\\
\label{vGamma}
\end{eqnarray}
where we have used the relations for $N_v$ and $\alpha (p)$ given in Eqs.~(\ref{ortho}) and (\ref{atrap}), respectively.

Now using Eq.~(\ref{w}) we obtain the functional form for the mode that contains information about the particle position, given by
\begin{equation}
{\bf w} ({\mbox{\boldmath$\Gamma$}}) = -i N_v \frac{\partial \overline{\bf E}_{\rm scat}^{{\rm I}+} ({\mbox{\boldmath$\Gamma$}},p) }{\partial p} \bigg |_{p=0}.
\label{wGamma}
\end{equation}
Note that this mode is only dependent on the scattered field.

We now calculate the SNR of the spatial homodyne and split detection schemes for particle sensing in an optical tweezers arrangement. Substituting the expressions obtained in Eqs. (\ref{atrap}) - (\ref{wGamma}) into Eq.~(\ref{optimalSNR}), the SNR for the spatial homodyne detection scheme is given by
\begin{eqnarray} \label{eqn:SNRSH}
{\rm SNR_{\rm SH,coh}} &=&  2 \alpha_w(p)\nonumber\\
&=&  2 \alpha(p) \left \langle {\bf w} ({\mbox{\boldmath$\Gamma$}},p), {\bf v} ({\mbox{\boldmath$\Gamma$}},p) \right \rangle\nonumber\\
&=&   - 2 i \sqrt{\frac{2\epsilon_0 V}{\hbar \omega}} \int^{\infty}_{-\infty} {{\bf w} ({\mbox{\boldmath$\Gamma$}},p)}^* \nonumber\\
& & \cdot \left ( \overline{\bf E}_{\rm scat}^{{\rm I}+} ({\mbox{\boldmath$\Gamma$}},p) + \overline{\bf E}^{{\rm I} +}_{\rm trap}({\mbox{\boldmath$\Gamma$}}) \right ) d {\mbox{\boldmath$\Gamma$}}\nonumber\\
&=& - 2 i \sqrt{\frac{2\epsilon_0 V}{\hbar \omega}} \int^{\infty}_{-\infty} {{\bf w} ({\mbox{\boldmath$\Gamma$}})}^* \cdot \overline{\bf E}_{\rm scat}^{{\rm I}+} ({\mbox{\boldmath$\Gamma$}},p)  d {\mbox{\boldmath$\Gamma$}}\nonumber\\
&=& - 2 K \sqrt{2 V} \int^{\infty}_{-\infty} {{\bf w} ({\mbox{\boldmath$\Gamma$}})}^* \cdot {\bf A } ({\mbox{\boldmath$\Gamma$}}) d {\mbox{\boldmath$\Gamma$}},
\end{eqnarray}
where the effective aperture function in the image plane co-ordinates is given by
\begin{eqnarray}
{\bf A} ({\mbox{\boldmath$\Gamma$}}) & = & {\cal F} \Big ( \sqrt{\frac{f_{\rm O}}{r_{\rm O}'}} \frac{e^{-ik (r_{\rm O}-r_{\rm O}'-f_{\rm O})}}{r_{\rm O}'} \Big[ \left  ( \hat {\bf r_{\rm O}}' \times \hat {\bf r_{\rm O}}' \times {\bf u}^{\rm O}_{00} \right )\hat {\bf l} \nonumber\\
& & + \left (\hat {\bf r_{\rm O}}' \times \hat {\bf r_{\rm O}}' \times {\bf u}^{\rm O}_{00} \right ) \hat {\bf n} \Big] \Pi_{R}({\mbox{\boldmath$\rho$}}) \Big ).
\end{eqnarray}

In a similar manner using Eq.~(\ref{snrSD}), the SNR for the split detection scheme is given by
\begin{equation} \label{eqn:SNRSD}
\mathrm{SNR_{SD, coh}} =  - 2 K \sqrt{2 V} \int^{\infty}_{-\infty} \mathbf{v}_f (\mathbf{\Gamma})^* \cdot {\bf A } (\mathbf{\Gamma}) d \mathbf{\Gamma}
\end{equation}
Correspondingly, the sensitivities for the spatial homodyne and split detection schemes can be conveniently calculated using Eqs.~(\ref{sensSH}) and (\ref{sensSD}), respectively.

\section{SIMULATION AND RESULTS}

A formal description for the trapping and scattered fields in an optical tweezers configuration was presented in Section~\ref{sec:opttweezer}. We now numerically solve for the scattered field from a particle trapped in the optical tweezers. We utilize the field imaging system shown in Fig.~\ref{schematic}~(b) to image the scattered field into a propagating optical beam that is subsequently detected. We compare the SNR and sensitivity of both split and spatial homodyne detection schemes (described in Section~\ref{sec:estimate}).

\begin{figure}[!ht]
\begin{center}
\includegraphics[width=8.5cm]{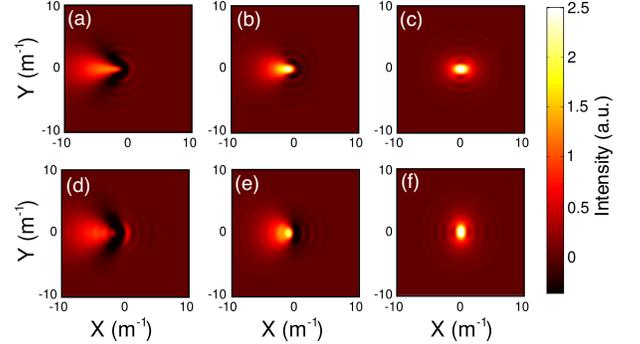}
\caption{Interference pattern of the trapping and forward scattered field in the far-field image plane for 200~mW trapping power, $\lambda = 1064$~nm, particle radius $a=0.1~\mu$m, permittivity of the medium $\epsilon_1=1$, permittivity of the particle $\epsilon_2 = 3.8$, and objectives with NA = 0.99 and focal spot size of $4~\mu$m. We assume absorptive losses in the sample are negligible. Figures (a)-(c) and (d)-(e) assume the trapping field is linearly $x$ and $y$-polarized, respectively. The color bar shows scale of the intensity distribution. The particle displacements are given by (a), (d): $1~\mu$m; (b), (e): $0.5~\mu$m; and (c), (f): $0~\mu$m.}
\label{fig:SDIntf}
\end{center}
\end{figure}

As mentioned in the preceding section, the origin of the co-ordinate system is defined to be at the focal point of the optical tweezers focusing lens system. The optical fields propagate in the $z$ direction and the scattering particle was assumed to be spherical and homogeneous. We model particle displacement in the $x$-$y$ plane, to illustrate the effect on the scattered field in the transverse plane. The far-field intensity distribution arriving at the detector is given by the interference between the trapping and forward scattered fields calculated from Eq.~(\ref{farfieldintensity}) and shown in Fig.~\ref{fig:SDIntf}. As the trapping field is far more intense than the scattered field, we have subtracted its intensity from the images shown in this figure as well as subsequent figures, to make visible the interference fringes between scattered and trapping fields.
Note that the terms due to just the scattered field have been ignored to reduce numerical error, justified since the total scattered power is four orders of magnitude smaller than the trapping beam power. The detection area was chosen to be larger than the area of the calculated image field to avoid inaccuracies due to clipping of the image. Notice that as the particle moves in one direction, the intensity distribution shifts in the opposite direction, due to the lensing effect of the objective. Note also the difference in intensity distribution between the $x$ and $y$ trapping beam polarization directions - i.e. the interference pattern appears ``compressed'' along the polarization axis due to the dipole scattering distribution of the particle.

\begin{figure}[!ht]
\begin{center}
\includegraphics[width=7.5cm]{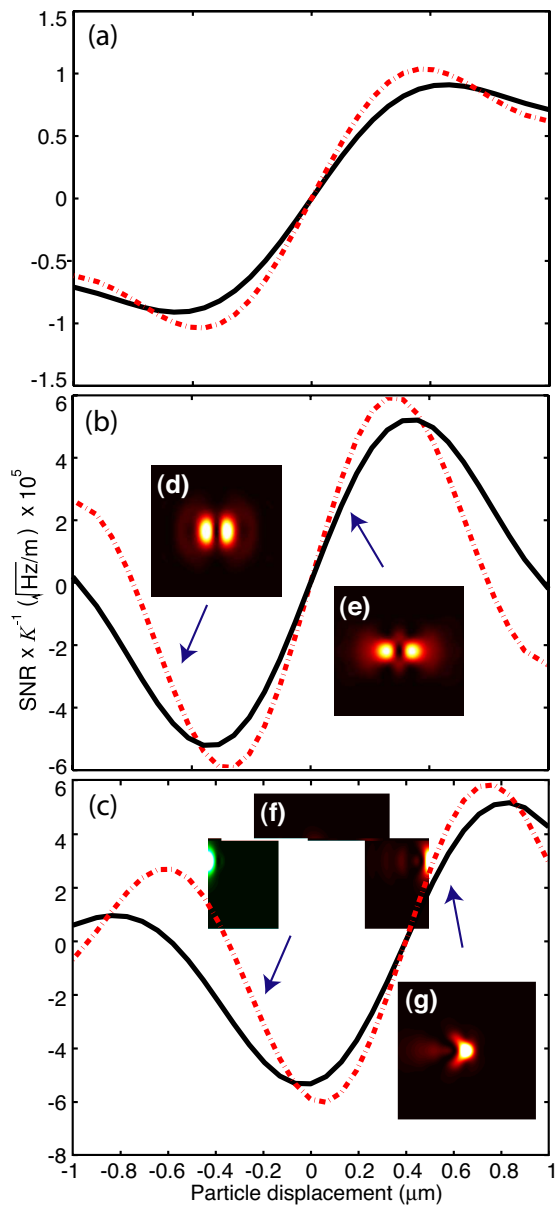}
\caption{Normalized SNR versus particle displacement for (a) split detection, (b) spatial homodyne detection with LO spatial mode optimized for small displacement measurements and (c) spatial homodyne detection with LO spatial mode optimized for larger displacement measurements. The black solid and red dashed lines are for linearly $x$ and $y$-polarized trapping fields, respectively. The LO spatial modes for the small displacement measurements are (d): $y$ and (e) $x$-polarized trapping fields, whilst for large displacement measurements are (f): $y$ and (g) $x$-polarized trapping fields.}
\label{SNRplot}
\end{center}
\end{figure}

The SNR for the split and spatial homodyne detection schemes were calculated, the results of which are shown in Figs.~\ref{SNRplot}~(a)-(c).
The SNR of the split detection scheme was evaluated by applying Eq.~(\ref{eqn:SNRSD}) to the calculated interference signal, shown in Fig.~\ref{SNRplot}~(a). To calculate the SNR for the spatial homodyne detection, the optimal LO mode first had to be determined. Improved SNR is possible with spatial homodyne detection when compared with split detection for all particle displacement regimes. However, the optimal LO mode depends on the position of the particle, so to achieve this a dynamical mode optimization routine would need to be implemented. Here we present results with detection optimized for two specific cases: (i) for a particle located close to the origin ($p \ll w_T$) as modeled in the theory section; and (ii) for a particle displaced from the origin by a factor of order $w_T$.

For the small displacement limit, the LO field was determined from the first order term in the Taylor expansion of Eq.~(\ref{w}) for the scattered field. The resulting SNR is shown in Fig.~\ref{SNRplot}~(b); with the corresponding LO spatial modes assuming $y$ and $x$ linearly polarized trapping fields shown in Figs.~\ref{SNRplot}~(d) and (e), respectively. One observes that for displacements significantly less than the trapping beam waist size the SNR is linear, with the optimum sensitivity - corresponding to maximum slope in the SNR -  occurring at zero displacement and significantly surpassing that achievable with split detection. Particle tracking with optimum sensitivity is possible in this linear regime. At particle displacements of around $\sim |0.4|~\mu$m, however, the SNR peaks. Small displacements of a particle around these points leave the SNR unchanged. Hence the signal read out from the spatial homodyne detector also remains unchanged, with the result that particle tracking becomes ineffective. As the particle position increases further, it moves out of the trapping field, causing a drop in the total scattered power and consequential exponential decay in the SNR.

It is possible to recalculate the LO field mode to optimize the sensitivity for particles fluctuating around any arbitrary position by performing a Taylor expansion in $p$ of the scattered field about that position, and retaining only the first order term. Fig.~\ref{SNRplot}~(c) shows the resulting SNR and corresponding LO mode shapes when the LO mode is optimized for particles fluctuating around 0.4~$\mu$m. Notice that now the maximum SNR slope, and hence optimum sensitivity, is shifted from zero displacement to displacements of around 0.4~$\mu$m. Hence, we see that as the tracked particle moves, it is possible to dynamically adjust the LO field shape to optimize the measurement sensitivity and hence the particle tracking.
\begin{figure}[!ht]
\begin{center}
\includegraphics[width=8.5cm]{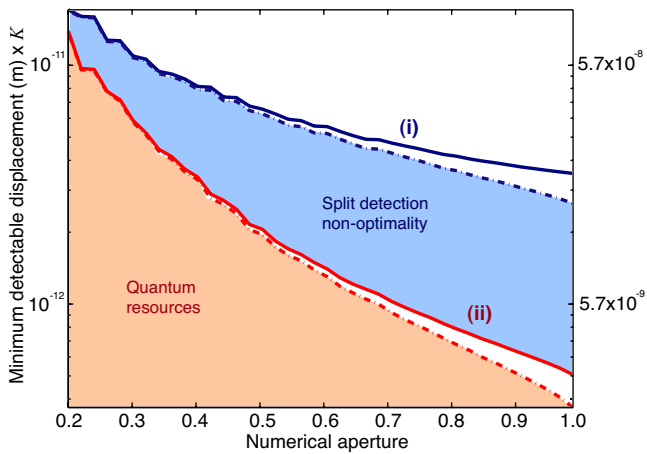}
\caption{Mininmum detectable displacement versus collection lens NA for (i) split and (ii) spatial homodyne detection, normalized by $K$. The solid and dashed lines are for linearly $x$- and $y$-polarized trapping fields, respectively. The axis on the right shows the minimum detectable displacement assuming 200~mW trapping power, $\lambda = 1064$~nm, particle radius $a=0.1~\mu$m, permittivity of the medium $\epsilon_1=1$, permittivity of the particle $\epsilon_2 = 3.8$, and objectives with focal spot size of $4~\mu$m. We assume absorptive losses in the sample are negligible. The {\it split detection non-optimality} shaded area shows the particle sensing sensitivity loss due to incomplete information detection from split detection. The {\it quantum resources} shaded area indicates the region where quantum resources such as squeezed light \cite{SpatialSQZ} can be used to further enhance the sensitivity of particle sensing measurements.}
\label{sensitivityplot}
\end{center}
\end{figure}

We now numerically evaluate the sensitivities of the split and spatial homodyne schemes in the small displacement limit, given by Eqs.~(\ref{sensSD}) and (\ref{sensSH}) respectively. The sensitivity is the minimum detectable displacement, defined as the displacement required to change the SNR by $1$. The respective sensitivity curves for (i) split and (ii) spatial homodyne detection versus the numerical aperture of the objective lens are shown in Fig.~\ref{sensitivityplot}.

The minimum detectable displacement for both the split and homodyne detection schemes decrease with increasing NA of the collection lens. As the NA increases, more of the scattered field is collected, therefore providing more information about the scattering particle. The spatial homodyne outperforms the split detection scheme for all NA values. This is due to the spatial homodyne scheme providing optimal information extraction of the detected field whereas the split detection scheme only measures partial information of the detected field, as derived in Eq.~(\ref{overlapSD}). Therefore curve (ii) is the quantum limit for particle sensing in optical tweezers systems. In order to perform measurements below this quantum limit, non-classical resources have to be used. For example, squeezed light in the spatial mode \cite{SpatialSQZ} corresponding to the displacement signal mode can be injected into the optical tweezers system to reduce the quantum noise floor and therefore enhance position sensing \cite{HsuJOB}.

\section{CONCLUSION}

We have developed a formalism for particle sensing in optical tweezers via the analysis of the transverse spatial modes imaged from a scattered field. The conventional quadrant detection scheme, used ubiquitously in optical tweezers experiments, was shown to only detect partial information from the scattered light field. We propose instead the use of spatial homodyne detection whereby optimal information from the scattering particle can be obtained via the appropriate transverse spatial mode-shaping of the LO field. A numerical simulation of the SNR and sensitivity of both split and spatial homodyne detection was presented and we demonstrate that up to an order of magnitude improvement in the sensitivity of spatial homodyne over split detection can be achieved.

\section*{ACKNOWLEDGMENTS}

This work was supported by the Royal Society of New Zealand Marsden Fund, and by the Australian Research Council Discovery Project DP0985078.


\end{document}